# Plasmonic Nanoslit Arrays Fabricated by Serial Bideposition: Optical and Surface-Enhanced Raman Scattering Study

*Samir Kumar, Yusuke Doi, Kyoko Namura, and Motofumi Suzuki*

Department of Micro Engineering, Graduate School of Engineering, Kyoto University, Nishikyo Kyoto 615-8540, JAPAN

**ABSTRACT** Recently, studies have been carried out on attempts to combine surface-enhanced Surface-enhanced Raman spectroscopy (SERS) substrates that can be based on either localized surface plasmon (LSP) or surface plasmon polaritons (SPP) structures. By combining these two systems, the drawbacks of each other can be solved. However, the manufacturing methods involved so far are sophisticated, labor-intensive, expensive, and also technically demanding. We propose a facile method for the fabrication of a flexible plasmonic nanoslit SERS sensor. We utilized the pattern on periodic optical disks (DVD-R) as a cheap substitute for printing the periodic pattern on PDMS with soft imprint lithography. Ag nanoslit (AgNS) was fabricated by serial bideposition using a dynamic oblique angle deposition (DOD) technique. The nanoslit structures were physically and optically characterized, and the experimental results were compared to the numerical simulation studies; Monte Carlo and the finite-difference time-domain (FDTD) simulation. The Ag nanoslit structure showed an excellent SERS enhancement, and its biosensing capability was demonstrated by the sensing of bilirubin.

# 1 INTRODUCTION

Surface-enhanced Raman scattering (SERS) is an enormous Raman signal enhancement in the presence of nanoscale features such as rough surfaces, wedge, sharp tips, metallic nanostructure edges, cracks or other discontinuities that allow ultrasensitive detection at a single molecular level.[1–3] SERS offers a wide range of applications across different fields, including biological sensing, trace analysis, medical diagnostics, forensics, pesticide detection, explosives, and medicines.[4–9] There are different methods to manufacture SERS substrates which can either be based on localized surface plasmon resonance (LSPR) or surface plasmon polariton (SPP).[10–12] The LSPR wavelength is dependent on the refraction index of the medium and the geometric shape, size, and structure of the nanoparticle. The coupling of between LSPR of the adjacent nanoparticles can enhance the SERS intensity by several order of magnitude.[13,14] However, in practice it is difficult to control the precise orientation and separation of nanoparticle. LSP-based SERS substrates suffer from the disadvantage of a complicated manufacturing method and low signal reproducibility.[15,16] SPP-based SERS substrate, such as a metal grating, can generate a more uniform electric field but has the disadvantage of low SERS signal enhancement.[17,18] Currently, there are a few studies on attempts to merge LSP and SPP structures to compensate for each other's disadvantages. By combining these two structures, a higher electric field concentration and a better SERS improvement can be achieved.[18–20]

Lithography, such as electron beam lithography, nanosphere lithography, or lithographic templates, is a well-known fabrication method for the generation of periodic plasmonic structure.

Shen et al. developed a strategy for SERS sensing based on grating-integrated gold nano-grasses (GIGNs). They used two-beam interference lithography to pattern a grating of photoresist bands, accompanied by molding with PDMS elastomer of its complementary structure.[21] They deposited the metallic layers by glancing angle deposition (GLAD) and tailored to obtain a nano-patterned surface of uniformly distributed SERS hotspots. Kalachyova et al. studied gold and silver nanoparticles of various shapes and dielectric functions immobilized onto the silver grating.[20] They reported that these structures could serve as dual-resonance SERS coupling systems between SPP-supported silver grating and LSP-excited grafted metal nanoparticles. They used a complex chemical process for Au and Ag nanoparticles synthesis and used a KrF excimer laser to produce a grating pattern on the photoresist. Xiao et al. demonstrated the polarization-independent performance of two-dimensional sinusoidal silver grating produced by maskless laser interference photolithography.[22] They demonstrated polarization-independent SERS performance of a 2D sinusoidal grating with enhancement factor ($E_f$) of around $10^5$ magnitudes. Wang et al. have reported a grating-like SERS substrate with tunable gaps based on Ag nanoislands on moth wing scales.[23] The grating-like substrate of rough and hierarchical Ag nanoislands exhibited EF around $4 \times 10^5$. Wang et al. reported a grating-like SERS substrate with tunable gaps based on Ag nanoislands on a moth wing scale. The manufacturing methods involved in all the studies were sophisticated, labor-intensive, expensive, and also technically demanding.

We report a cheap and facile method for the fabrication of nanograting coupled SERS substrate by serial bideposition of Ag on periodic optical disks (DVD-R) patterned PDMS. DVD has a periodically structured groove with a periodicity of about 750 nm. We used this DVD pattern as a cheap substitute for printing the periodic pattern on PDMS with soft imprint lithography. The Ag nanoslit (AgNS) SERS sensor was fabricated by serial bideposition of Ag using dynamic oblique angle deposition (DOD) technique. We report the experimental results and numerical simulation studies on the growth and optical properties of the AgNS SERS sensor. The nanoslits structure showed an excellent SERS enhancement, and we were able to detect biomarker *bilirubin* using the AgNS SERS sensor with intensity counts even better than the commercially available SERS chip Wavelet from Nidek, Japan

## 2 MATERIALS AND METHODS

### 2.1 Transfer of periodic DVD pattern

Flexible PDMS surface was patterned using DVD-R (Scheme S1, supplementary information). DVD-R was used as a master for transferring the periodic pattern onto the PDMS. The DVD-R is cheap and readily available. The steps followed to prepare the patterned polycarbonate surface from the DVD-R is given in detail as supporting information. The steps followed for the transfer of DVD-R pattern onto PDMS is shown in Scheme 1. DVD-R has a periodic nanoscale groove. The periodicity ($\lambda$) and amplitude ($A$) of the grating pattern (750 and 100 nm, respectively), shown in Scheme (e), were calculated from the AFM images using gwyddion software(v2.53).[24] The PDMS was prepared by mixing prepolymer (Dow Corning Sylgard 184) with a curing agent in a ratio of 10:1 at room temperature. Next, the air bubbles of the mixture were removed using a rotary pump and a desiccator. After that, the mixed PDMS solution from which bubbles were removed was poured onto the DVD-R and was cured for 30 minutes at 75 °C. After drying the PDMS was peeled off the DVD-R template to obtain a patterned PDMS substrate.

### 2.2 Deposition of Ag on the patterned PDMS substrate

DOD was used to deposit AgNS on the patterned PDMS substrates by serial bideposition. DOD is a versatile, simple, reproducible, and low-cost technique for the fabrication of metal and metal oxide nanostructures for SERS and photocatalysis applications.[25–29] It is a physical vapor deposition technique where the vapor flux is incident at a large angle (α = 85 °) with reference to the normal surface to the substrate. Nanostructures such as nano-helix and nanosprings can also be fabricated using DOD by rotating the substrates azimuthally.[30,31] A schematic view of the steps for the deposition of *Ag* by DOD is shown in Scheme 2. Four AgNS samples were grown by DOD with thicknesses 200 nm, 400 nm, 600 nm, and 1000 nm (thickness shown in QCM) at an angle of 76⁰ with respect to the substrate normal. Henceforth, these samples will be referred to as S2, S4, S6, and S10, respectively. All four AgNS samples were fabricated using an electron-beam evaporation system at a deposition pressure of 6×10⁻⁵ Pa. Ag source material was deposited onto the patterned PDMS at an angle of 76° at a deposition rate of 3-5 Å/s as measured by a 6 MHz quartz crystal thickness monitor (QCM).

The deposition angle was chosen as 76 ° to make the shadow region large enough to create nanoslits. Also, the flux of incident Ag adatoms was perpendicular to the grating pattern to deposit Ag only along the ridges and not within the trench of the periodic structure. To improve the uniformity of the deposited Ag material, the in-plane deposition angle was periodically rotated by 180 ° each time the QCM thickness increased by 5 nm; that is when the average silver deposited film thickness on the sample increased by 1.2 nm. For calculating the deposition angle, we assumed that the surface normal at the peak of the crest would make and angle $\alpha$ with the incident flux and *x* is the required width of the slit then the deposition angle

$$\alpha = tan^{-1}\left(\frac{375 + \frac{x}{2}}{100}\right) \quad (1)$$

The angle of deposition was selected, such that to obtain a slit width of approximately 100 nm.

**2.3 Optical and SERS measurements**

The optical properties of the AgNS samples were investigated using transmission measurements at normal incidence. In the ultraviolet light region, a deuterium light source (DH-2000, Ocean Optics) was used, and in the visible light region, a halogen light source (LS-1, Ocean Optics) was used for transmission measurements. The light guided by the optical fiber was collimated by the convex lens and was incident on the AgNS sample perpendicular to the sample surface. Transmission spectra were collected using a spectrometer (Ocean Optics USB4000).

SERS spectra were obtained using a Raman spectrometer (RAM200, LambdaVision). A 785 nm laser with an x50 objective and 15 mW power, on the sample, was used for the excitation purpose. The spectral acquisition time was 1 s. SERS measurements were performed at five random spots on the substrate. All the measurements performed at room temperature. Rhodamine 6G (Rh6G, 99.9 + %, Sigma Aldrich) and *trans*-1,2-bis(4-pyridyl) ethylene (BPY, 99.9+%, Sigma) were used as a Raman probe in this study. Rh6G and BPY solutions were prepared in Milli-Q water. A 10 μL droplet of Rh6G and BPY solutions was released onto the AgNR substrates and allowed to dry prior to data acquisition.

**2.4 Surface morphology**

A scanning electron microscope (SEM; Hitachi High Tech. SU3800) with a LaB6 detector in the secondary electron mode operating at an acceleration voltage of 10 kV was used to conduct surface morphology analysis of as-prepared samples. The surface morphology was obtained across 3 μm × 3 μm area using atomic force microscopy (AFM; Hitachi AFM5000II) in tapping mode.

**2.5 Modelling and Simulation of growth on a periodic substrate**

The Virtual Film Growth System (VFIGS) based on the Monte Carlo (MC) ballistic deposition (BD) process were used to simulate the growth process and reproduce the morphology of the AgNS film on a periodic surface.[32] The VFIGS describes thin films as an aggregate of cubic particles. Particles are added to a growing structure with randomly generated positions and directions. The deposition is started from randomly selected sites in the *x-y* plane and just above the growing film surface. The hopping direction is determined by the deposition conditions such as the deposition angle $\alpha(=76°)$, the dispersion of the deposition angle $\Delta\alpha(=0.5°)$, and the azimuthal deposition direction $\phi(=180°)$. The number of steps in a random walk, $S(=10)$, is related to the diffusion length. The hopping probability from a site *i* to an unoccupied adjacent site *j* is defined as

$$P_{i \to j} = \frac{exp\,(\gamma N_j)}{\sum_j exp\,(\gamma N_j)}, \qquad (2)$$

where $N_j$ is the number of the site *j* with the adjacent particles, $\gamma$ is a constant, and the summation is taken over all the permitted sites. Physically the constant $\gamma$ represents the ratio of surface energy to temperature.[32]

We unit of the length is defined as ''*u*'' where 1 *u* is equal to the length of the edge of each cube. $L(=800\,u)$, $M(=800\,u)$, and $N(=600\,u)$, *M*, and *N* cells exist in the *x*, *y*, and *z* directions, respectively. A particle occupies one of the cubic cells within this volume. The sample is deposited on a periodic substrate with trench height $H=14$ u and thickness 3 *u* until the maximum height of the surface reaches *3.6 u*.

## 2.6 Finite-Difference Time-Domain (FDTD) Simulation

FDTD is an algorithm for time propagation which incorporates Maxwell's curl equations into the time domain. The electric field is determined at any arbitrary geometry and time of the simulation. Three-dimensional FDTD was performed using a commercial software package (Lumerical FDTD: 3D Electromagnetic Simulator), to study optical transmission and electric field strength of the AgNS samples.[33] The simulations were conducted with two mutually orthogonal polarizations called "*s*" (light polarized parallel to the long nanoslit axis) and "*p*" (light is polarized perpendicular to the long nanoslit axis). We used a long simulation time of $1000\,fs$ to achieve the Courant stability of 0.99 with a step of $0.018\,fs$ to ensures the stability for the plane wave excitation. In our simulation, we set the auto-cut value to $1 \times 10^{-5}$, which can be further reduced. The periodic boundary condition is used to extend the geometry to infinity in both x and y- directions. Field components in ±z directions at grid edges are absorbed by perfectly matched layers (PMLs).

## 3 RESULTS AND DISCUSSION

**3.1 Fabrication and morphology of the nanoslit films**

The SEM (Fig. 1) and AFM images (Fig S1, Supplementary Information) images show that a regular periodic nanoslit pattern is formed on the surface of the PDMS. The individual sub-deposits were so thin for these depositions that they did not grow like zigzag nanostructures. The directional growth of Ag nanostructure occurs in the direction parallel to the plane of deposition and perpendicular to the pattern of the substrate. The morphology of the Ag on patterned PDMS is shown in Fig. 1(a)-1(d). We can derive the normal thickness of the Ag deposited ($h_0$) on the periodic surface from simple geometrical arguments as

$$h = h_0 \times \cos 76. \qquad (3)$$

The details of measured height and roughness from AFM images, slit width, and film width from SEM images, and calculated height from Eq. 3 is given in Table 1. As can be seen, the calculated height corresponds very well to the average height $h$ obtained from the AFM. The height (≈ 270 nm) calculated from the cross-section SEM of the sample S10 also validates our calculation of the film thickness.

*Table 1*. Calculated height and roughness from AFM images; slit width, and film width from SEM images, and calculated height

| Sample | QCM Thickness, $t$ (nm) | Film thickness, $h$ (nm) (calculated) | Slit width, $s$ (nm) | $h/s$ | LSPR (nm) | SPP (nm) | Roughness | $h$ (from AFM) |
|---|---|---|---|---|---|---|---|---|
| S2 | 206 | 50 | 170.6 | 0.29 | 720 | 754 | 39.2 | 53.5 |
| S4 | 413 | 100 | 149.3 | 0.67 | 713 | 760 | 42.5 | 97.4 |
| S6 | 620 | 150 | 115.5 | 1.3 | 704 | 767 | 39.6 | 148.2 |
| S10 | 1033 | 250 | 74.5 | 3.36 | 685 | 790 | 34.2 | 260.1 |

The cross-sectional SEM image of the sample S10 is shown in Fig. 1(e). The film at the top of the crest and its edges were found to be slightly thicker. Since the substrate has a periodic non-uniform surface, it is difficult to estimate the growth direction of the Ag nanostructure as it will change at every point on the periodic pattern. However, we can make a rough estimate of the film's growth trajectory and understand the growth mechanism by using the empirical tangent or cosine law of thin films deposited at slightly different incident angles. In general, the column tilt angle ($\beta$) is less than the incident angle of the vapor flux and follows the empirical cosine rule for $\alpha > 60^o$,[26]

$$\beta = \alpha - \sin^{-1}\left(\frac{1-\cos(\alpha)}{2}\right) \qquad (4)$$

Since the normal to the surface of the periodic patterned PDMS structure is spatially variable, the material growth rate and the surface nucleation site of the periodic PDMS structure will also be spatially variable, resulting in a non-uniform film thickness along the ridge. The flux direction (orange dotted lines), the columnar growth direction (gray arrows), and the film thickness showing the net deposited film thickness (black shaded region) at different points of the periodic substrate measured using Eq. 4 are shown in Fig. 2. The growth direction will be non-uniform at the top of the periodic peaks and along the edges of the pattern due to the spatial variability in the sticking coefficient of the adatoms.

The AgNS samples have a primary periodic grating structure with a shadowed plateau and nanoslits. It was found that nanoslit widths decrease by increasing Ag film thickness, Fig. 1(g). The slit width decreases monotonically from 175 nm up to 75 nm for film thickness 50 nm to 250 nm. Thus, the slit width, $s$, is inversely related to film thickness, $h$. By increasing the film thickness, we can tune the aspect ratio $h/s$ ratio from 3.4 to 0.3. We can also see that the columnar growth becomes more prominent as the thickness of the silver increases. Due to the large deposition angle, the Ag decorate only the surface directly facing the flux, separated by periodic gaps in the shadowed regions. The roughness values obtained from AFM images in root-mean-square (RMS) were 28.7, 39.2, 42.5, 39.6, and 34.2 nm for bare DVD-R patterned PDMS, S2, S4, S6, and S10 samples, respectively. Initially, the Ag nucleation sites increase the surface roughness compared to the bare patterned PDMS surface. The roughness of the AgNS sample increases until the growth of 100 nm Ag after that; it starts decreasing. This reduction in surface roughness may be attributed to the coalescence of the nano-island and the columnar growth may become more pronounced. The SEM images (Figure 1(a)-1(d)) were statistically analyzed through the ImageJ software suite to evaluate the average nano-island transverse width ($p_{\parallel}$) and length ($p_{\perp}$) projected on the horizontal plane (Fig. S2 supplementary information). With an increase in film thickness, the projected width and length and of the nano-islands increase linearly. Another point to note is that the $p_{\perp}/p_{\parallel}$ ratio increases with the film thickness. This may be due to the preferential diffusion along one direction and dominance of the shadowed columnar growth.

As already discussed above, the periodic pattern on the PDMS substrate surface has an inclination, so that at each point on the periodic pattern the surface normal has a different angle to incident angle of the adatom flux. The interplay between the ballistic shadowing (which produces columnar structure) and surface diffusion (results in uniform coverage) results in the final morphology of the nano grating surface.[34]

For better visualization of the growth process, growth on the periodic pattern was simulated using the VFIGS 3D Monte Carlo ballistic code.[32] VFIGS was used to simulate the start of shadowing-dominated growth in films deposited in glancing angles.[35] A cross-sectional and top view of a typical growth at $\alpha = 76°$ generated by VFIGS is shown in Fig. 2., where $t = 0$ and $t = T$ represents the start and end of growth, respectively. The cross-section and top view images generated in Fig. 2(b) resembles the SEM images Fig. 2(c) and the growth model is remarkably consistent with our interpretation of film growth on the periodic surface. Growth begins with the growth of several nano-islands, mostly on the top and edges of the crest of the periodic pattern. Shadowing effect dominates growth in the initial stages. The shadowing effect, along with the low diffusivity of the adatoms, alters the mode of thin film growth on the periodic surface. The growth increases quite sharply after the formation of the initial shadowing centers. Since the angle of incident flux is different for the top and edges of the crest, the direction of growth of the columns is different as well. The columns grow in the vertical direction at the top of the crest, while at the edges, they grow towards the direction of the flow in the edges. The lateral growth of the columns on the edges of the crest accounts for the reduction of the nanoslit width with the increase of the thickness of the film. We can, therefore, conclude that the nano-slit width can be controlled by adjusting the deposited film's thickness at a certain angle.

**3.2 Optical Properties of Flexible AgNS SERS Substrate**

Metal nanoslit arrays are ideal for plasmonics applications because of their ability to combine far-field electromagnetic radiation with electromagnetic surface modes. The confined plasmon surface modes on the interface contribute to an increased magnetic field.[17] The optical transmission measurement was performed to determine the LSPR position of the nanoslit samples. The experimentally observed transmission spectra of s-polarized (light polarized parallel to the nanoslit long axis) and p-polarized (light is polarized perpendicular to the nanoslit long axis) for AgNS samples is shown in Fig 3(a). The transmission spectra for the samples generally consist of one peak around 320 nm (not shown in the figure) and three dips around 560 nm, 725 nm, and 750 nm. The peak around 320 nm corresponds to the bulk plasmon mode of the Ag. The transmission dip of around 720 nm and 560 nm corresponds to the SPP mode, as confirmed by the FDTD simulation. When the thickness of the Ag layer increases from 50 to 250 nm, the 750 nm SPP mode is slightly red-shifted due to changes in the periodic structure (Fig S4, supplementary information). In the case of *s*-polarized light, the electrons are free to oscillate unrestricted along the connected metallic nanostructures, and the optical spectrum is close to that of a continuous film.[36] For p-polarized light, electron confinement effects become significant, and an LSPR at around 685-720 nm is visible in the spectrum. The particle LSPR wavelength $\lambda_{lsp}$ can be obtained using the Drude model of the metal as

$$\lambda_{lsp} = \lambda_p \sqrt{(2n_d^2 + 1)} \approx \sqrt{3}\lambda_p \left[1 + \frac{2}{3}(n_d - 1)\right] \quad (5)$$

where $\lambda_p$ is the bulk plasmon wavelength, and $n_d$ is the refractive index of the surrounding medium. Equation (5) predicts an LSPR of wavelength 712 nm, which is within the experimentally observed value of around 685-720 nm for different samples.

The difference in the experimentally measured LSPR magnitude can be because the nanoparticle size plays a significant role in determining the wavelength of the plasmon resonance. The optical characterization results were contrasted with those of the FDTD simulations. The transmission for sample S4 obtained from FDTD simulation is shown in Figure 3(b). The inset shows the AgNS structure used for the simulation process. The experimental and FDTD transmission are in good agreement. The optical transmission obtained by FDTD for all AgNS samples is shown in Fig. S5 (Supplementary Information). We observed only one sharp dip at around 750 nm due to SPP for FDTD simulation. The dip due to LSPR was not observed in the simulated result as we considered the AgNS film to be a uniform Ag thin film and not an accumulation of nanoparticles. For a fixed structural period, the SPP resonance wavelength remains approximately constant, which is also confirmed by the FDTD simulation. In the case of experimental transmittance, the dip position is slightly shifted from 750 nm which may be due to the irregularity in the periodicity during pattern transfer. Experimentally, we observed a slight redshift in the SPP resonance with increasing film thickness, which can be attributed to a change in the ratio of $s/h$ (slit width/thickness) from 3.4 to 0.3 with an increase in the film thickness. For LSPR, various factors contribute to the shift in wavelength. Firstly, the refractive index of the surrounding; secondly, the distance between the two nanoparticles; and thirdly, the size and shape of the metal nanoparticle plays a significant role in deciding the plasmon resonance wavelength. The role of the refractive index on changing the resonance peak may be negated since both the medium and the substrate were the same for all samples. Increasing the distance between the nanoparticles results in a small blue shift of the resonance peak. The LSPR is sensitive to nanoparticle shape changes in in-plane and out-of-plane directions.[37] Increasing the in-plane nanoparticle width results in a blue shift of the wavelengths. Increasing the out-of-plane height (by increasing the film thickness) results in a redshift of the LSPR wavelengths. The trade-off between the increasing distance of the particles, the increase of the in-plane and the out-of-plane height of the nanoparticles results in a shift of the LSPR as a whole.[37] The change in the in-plane height and nanoparticle separation seems to be dominant because of the blue shift of the resonance with the film thickness. For our AgNS samples, the ratio of the projected length/width of the nanoparticles increases more than twice with the increasing thickness (Fig. S2, supplementary information). Consequently, the change in the in-plane height and/or separation of nanoparticles tends to be dominant due to the blue shift of the resonance with the film thickness.

The optical transmission through the AgNS samples was compared to the transmission through a uniform film of the same thickness with unpolarized light, Fig S6. (supplementary information). The transmission of the film decreases with increasing Ag thickness. Around the bulk plasmon wavelength, $\lambda_p$ ($\approx$ 320 nm), the film significantly transparent. The AgNS samples showed an enhanced transmission compared to the continuous film. The larger transmission was observed at longer wavelengths ($\geq$ 600 nm) compared to continuous thin films.

It should be noted that the resonance wavelength of both LSPR and SPP varies between 680-790 nm. It is important to enhance both the incident and the scattered fields in order to maximize the SERS enhancement factor.[38] Improvement of both fields is optimally achieved when incident frequencies and scattered fields span the LSPR extinction spectrum. Thus, by choosing a laser with an appropriate wavelength, the SPP and LSPR can be coupled to the excitation wavelength, which can lead to a significant improvement of the Raman signal.

**3.3 SERS Studies on Flexible AgNS Substrates**

The average SERS spectra of Rh6G and BPY on AgNS substrates are shown in Fig. 4(a) and 4(b), respectively. For clarity, the spectra were plotted by offset in the increasing order of their SERS intensity. For Rh6G, the Raman bands at 613, 775, 1190 cm$^{-1}$ are due to $C-C-C$ ring in-plane, out-of-plane bending, and $C-C$ stretching vibrations, respectively. The other Raman bands at 1364, 1508 and 1646 cm$^{-1}$ are assigned to aromatic $C-C$ stretching vibrations of Rh6G molecule.[39] The SERS spectra of BPY has four major main bands at around 1015, 1234, 1294 and 1610 cm$^{-1}$ is due to the pyridine ring breathing, ring deformation, $C=C$ in-plane ring mode, and $C=C$ stretching mode, respectively.[40,41]

Figure 4 clearly shows a significant improvement in the Rh6G and BPY SERS signals by the AgNS substrate. The SERS response of the nanoslit samples was found to be comparable to that of commercially available SERS chip Wavelet$^{TM}$ from Nidek, Japan. The SERS intensity for the sample S4 was even better than the Wavelet$^{TM,}$ which demonstrates its potential as a SERS substratum. The average SERS performance of the four samples first increased with an increase in the thickness of up to 100 nm, then decreased and increased again for the thickness of 250 nm. The SERS performance for sample S10 was still lower than for sample S4. The best SERS enhancement of sample S4 may be due to its highest roughness among the four samples (Table 1). When a uniform electric field is applied to a rough conductor, the electric field near the surface is not uniform but is largest near the sharpest surface feature.[2] The roughness of the substrate is, therefore, a crucial factor that contributes to the improvement of SERS by the lightning rod effect.[42] Sample S6 has higher roughness than sample S10 but the later one showed better SERS enhancement. In addition to roughness, the thickness of the metallic layer plays a significant role in the SERS activity.[43] The thickness of the sample S10 is higher than that of sample S6 and therefore, the thickness factor may be a reason for better SERS enhancement of sample S10. The thickness factor also explains the almost comparable SERS performance of Sample S10 with respect to Sample S4, which has a large roughness compared to the former.

The enhancement factor ($E_f$) was calculated to evaluate the SERS performances of our AgNS SERS substrate. The $E_f$s are difficult to compare because of the difference in experimental conditions. There are primarily two types of $E_f$s; the maximum $E_f$ and the average $E_f$.[44] The maximum $E_f$, is rarely used in practice because it is difficult to locate the sites for the maximum SERS enhancement called hot-spots. The analytical SERS $E_f$ is the average increase in the scattering volume and is more reproducible. The analytical enhancement factor can then be defined as

$$E_f = \frac{\frac{I_{SERS}}{C_{SERS}}}{\frac{I_{RS}}{C_{RS}}} \quad (7)$$

where an analyte solution with concentration $C_{RS}$, produces a Raman signal $I_{RS}$ under non-SERS conditions on conventional Ag thin film.[44] The background-corrected peak height of the Rh6G peak at around 1364 cm$^{-1}$ was used to measure the overall response to SERS. Using equation (7), the $E_f$ for the nanoslit samples was calculated using traditional Ag thin film as reference. The $E_f$ was calculated to be ~10$^5$ for the 1364 cm$^{-1}$ peak.

Sensitivity and consistency are the foremost concerns for all SERS substrates. Sample S4 was used for the sensitivity and reproducibility test as it showed the maximum SERS enhancement. We were able to detect Rh6G up to a concentration of 5 x 10$^{-7}$ g/ml on the S4 sample. SERS measurements were carried out at 10 different locations on the same sample to evaluate the reproducibility of Rh6G SERS signals. The intensity of the 1364 cm$^{-1}$ peak for 10$^{-6}$ M Rh6G solution from 10 different sites on the nanoslit substrate is shown in Fig. S7, supplementary information. The standard deviation of the intensity on the sample was found to be less than 10%.

To evaluate the local field enhancement in nanoslit substrates, we analyzed the electric field distribution using FDTD simulations. The structure used for our FDTD simulation and its corresponding local field enhancements ($|E_{local}/E_0|$) are shown in Fig. 5(a) and 5(b), respectively. Our geometries for the simulation are an approximation of the actual geometries shown in the SEM images. 785 nm wavelength was used for the FDTD simulations. The local field enhancement for the sample S4 was calculated as it was found to have the highest SERS $E_f$. In our simulation, we did not include the effect of nanostructures on the optical properties, as we were interested in the effect of nanoslit. The $E$-field confinement is particularly strong around the edges of the AgNS, where the $E$-field enhancement is the highest. The higher plasmonic activity of the AgNS due to the sharp edge on the nanoslit is noticeable. Hence, the creation of hot spots on the nanoslit edges may be responsible for the significant improvement in the Raman signal of Rh6G molecules. The optical transmittance was also calculated using the FDTD, Fig. S5 (supplementary information). The thickness of the Ag film was varied from 50 nm to 250 nm with corresponding slit width varying from 175 to 75 respectively, as observed experimentally. As discussed above, the optical transmission is in good agreement with the experimental observations.

We have demonstrated the potential application of AgNS SERS substrates through bilirubin detection. Bilirubin is a tetrapyrrole pigment, and a toxic metabolite of heme reduced excretion of which results in jaundice, a well noticeable symptom of liver disease. Hyperbilirubinemia is a condition of higher-than-normal levels of bilirubin in the blood. For adults, this is any level above 0.1 mg/ml and for newborns 0.2 mg/ml and critical 0.25 mg/ml. The SERS spectra of .06 mg/ml bilirubin solution, which is much below the normal level, on AgNS sample S4, is shown in Fig. 6.

The SERS spectra of bilirubin 7 major main bands at around 1613, 1450, 1257, 1045, 986, 687, and 642 cm$^{-1}$ can be assigned to $C=C$ the ring stretching, $C-N/C-C$ stretching, $C-H$ wagging, $C-N$ stretching, $CH_3$ deformation, $C=O$ out of plane bending, and methylene bridge deformation, respectively.[45] The detailed assignment of the various Raman bands for bilirubin is given in Table S1 (supplementary information). The identification of significant bilirubin bands at concentrations well below the normal level illustrates the sensing capacity of these AgNS SERS substrates.

## 4 CONCLUSIONS

In conclusion, we report the fabrication of a sensitive and facile SERS sensor using commercial DVD-R optical discs as nanoslits-based flexible polymer substrates. Serial bideposition by dynamic oblique angle deposition can be used to produce nanoslit SERS substrates reproducibly with varying slit width. The highest enhancement was detected on Ag nanoslit samples with a thickness of 100 nm. The transmission spectra confirmed the excitation of both SPP and LSPR by the nanoslit sensors. The label-free biosensing capability of the proposed Ag nanoslit plasmonic structure was evaluated by real-time detection of biomarker bilirubin well below the normal concentration.

## ACKNOWLEDGMENTS

This work was supported by the grant Center of Innovation Program (COI) from the Japan Science and Technology Agency (JST), "The Last 5X Innovation R&D Center for a Smart, Happy, and Resilient Society" (grant number JPMJCE1307).

## ASSOCIATED CONTENT

**Supporting Information**

Schematic showing the construction of grooved polycarbonate DVD grating substrate (Scheme S1), The AFM images of the AgNS film (Figure S2), Experimental and simulated transmission spectra for the AgNS samples and comparison with thin film (Figure S3-S6), reproducibility test of sample S4 (Figure S7), Table S1 listing the Raman shift bands of bilirubin.

## AUTHOR INFORMATION

**Corresponding Author**

*E-mail: samir.kumar.2r@kyoto-u.ac.jp

# Plasmonic Nanoslit Arrays Fabricated by Serial Bideposition: Optical and Surface-Enhanced Raman Scattering Study

*Samir Kumar, Yusuke Doi, Kyoko Namura, and Motofumi Suzuki*

Department of Micro Engineering, Graduate School of Engineering, Kyoto University, Nishikyo Kyoto 615-8540, JAPAN

# Figures

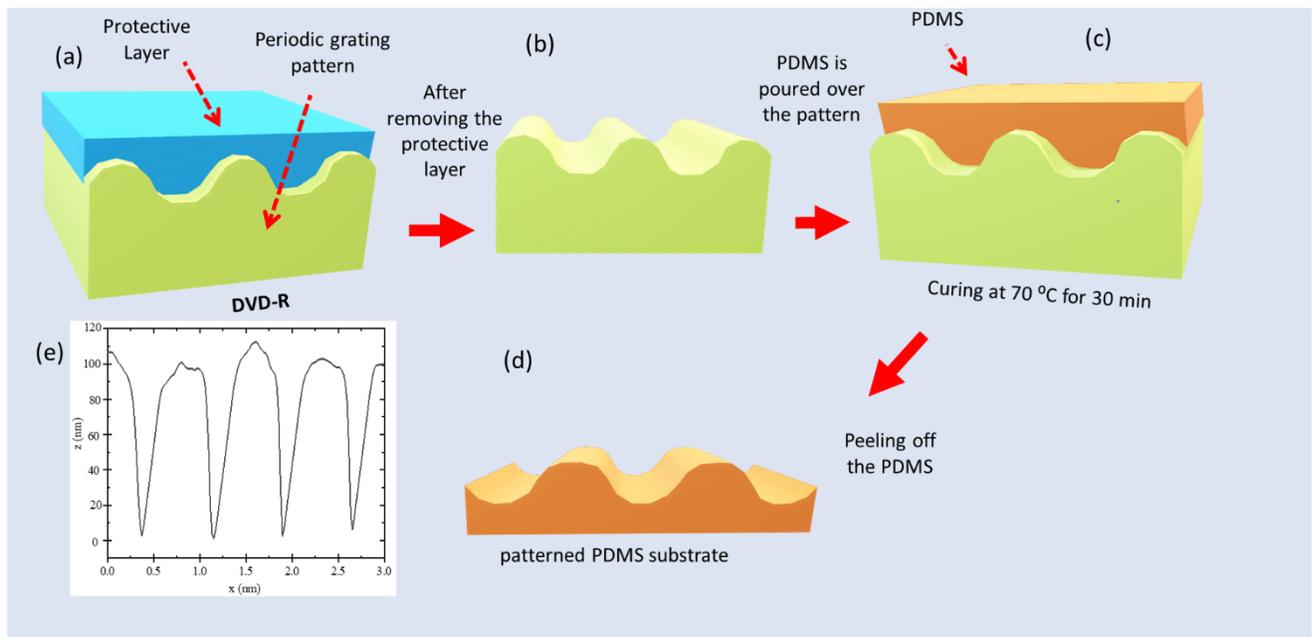

**Scheme 1.** Schematic diagram of the steps followed for the pattern transfer procedure onto PDMS. (a) Cross sectional view of the DVD-R. The periodic structure of the DVD-R disc is protected by a polycarbonate layer; (b) protective layer was removed to expose the periodic structure using a sharp blade; (c) liquid PDMS was poured on the exposed DVD-R pattern; (d) the DVD-R with PDMS was cured in the oven for about 30 min at 75 ⁰C; (e) the cured PDMS was peeled off from the DVD-R surface to obtain the patterned PDMS substrate. (e) AFM line profile of the patterned PDMS obtained after pattern transfer.

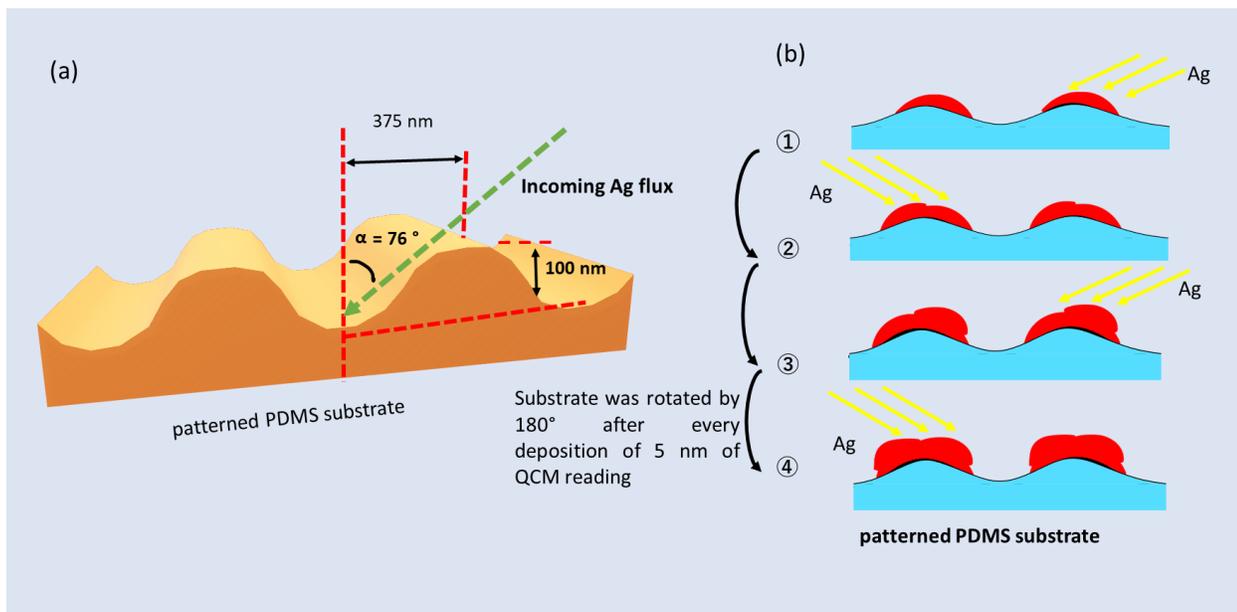

**Scheme 2** A schematic view of the steps for the deposition of Ag by DOD. (a) Shows the direction of Ag flux with respect to the periodic pattern. After the transfer of pattern, Ag was deposited on the PDMS substrate by DOD. The Ag flux was incident at an angle of 76⁰ to the surface normal and perpendicular to the grating pattern.; (b)sss After the deposition of 5 nm of Ag, the sample was rotated by 180⁰ to create a nanoslit in the valley of the grating due to shadowing effect. Films of thicknesses 50 nm -400 nm were deposited by this method.

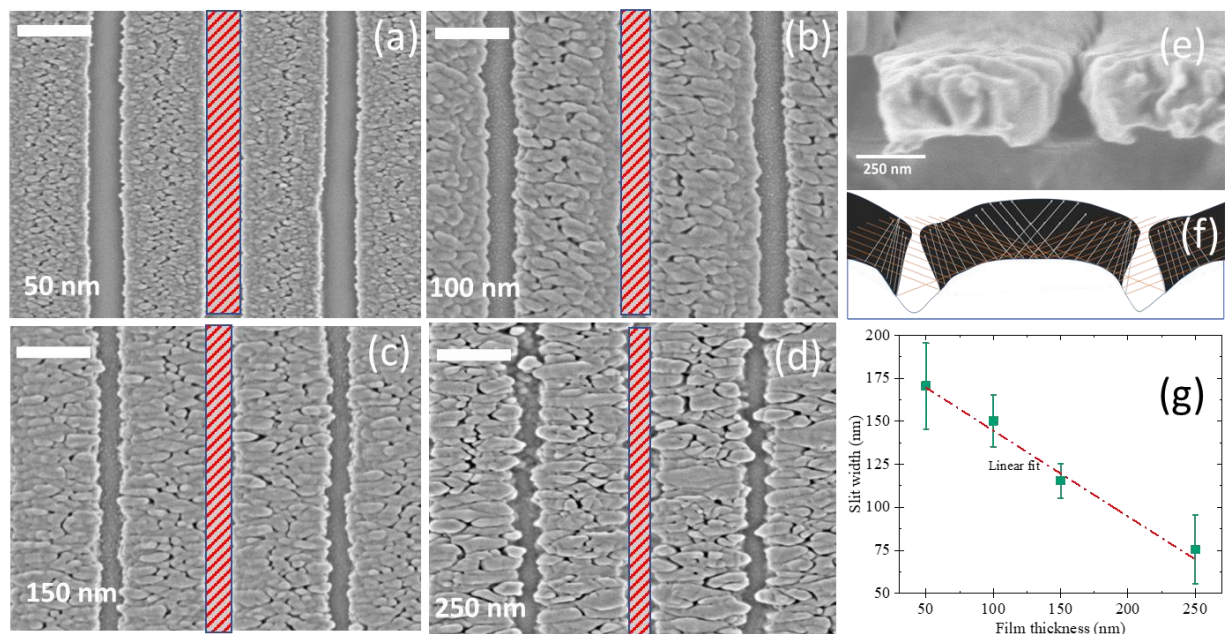

**Figure 1** SEM images (top view) of the Ag nanoslit samples with (a) 50 nm, (b) 100 nm, (c) 150 nm, and (d) 250 nm thickness; (e) Cross-sectional SEM of sample with thickness 100 nm; (f) show the incident flux direction (orange dotted lines), the columnar growth direction (grey arrows) and the film thickness showing the net thickness of the deposited film (black shaded region) at different point of the periodic substrate. (g) slit width as a function of deposited film thickness. Scale bar corresponds to 250 nm.

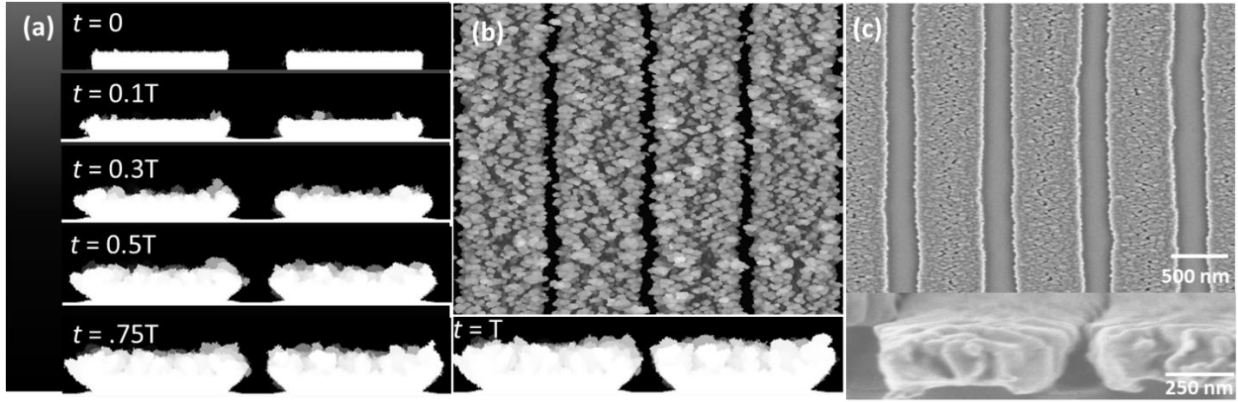

**Figure 2** (a) Shows the cross-sectional view of the time evolution of AgNS growth obtained by Virtual Film Growth System, where t=0 and t= T represents the start and end of growth respectively; (b) Simulated top and cross-sectional view of AgNS film at t=T; (c) Top and cross-sectional view of experimentally grown AgNS sample. The generated cross-sectional and top view images in Fig. 3(b) resembles the scanning electron micrograph SEM image Fig. 3(c), and the growth model agrees remarkably well with our interpretation of film growth on the periodic surface.

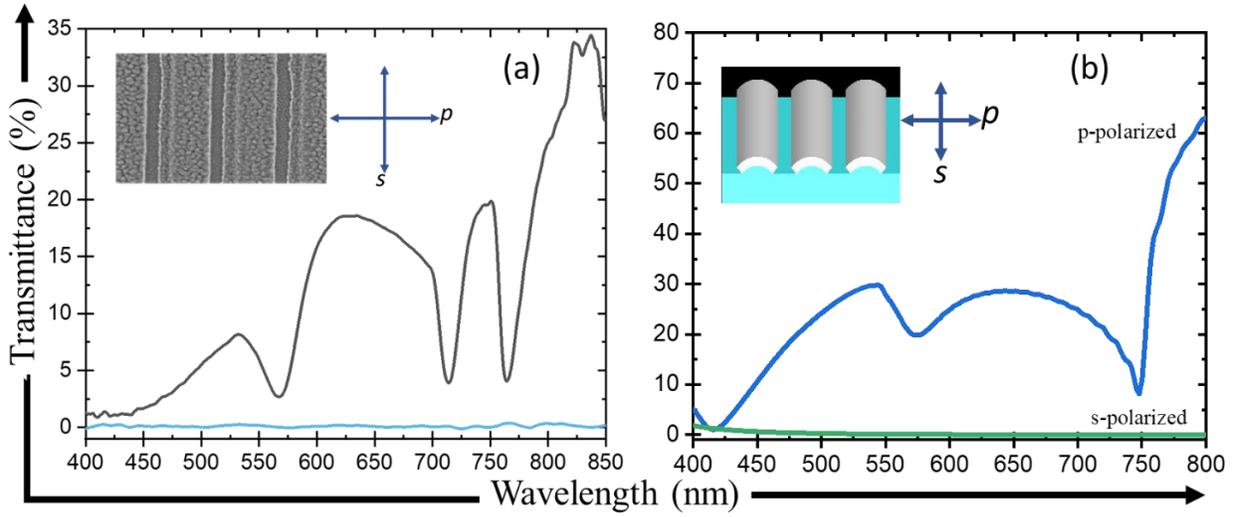

**Figure 3**. (a) Experimental transmission spectra, and (b) transmission obtained from FDTD simulation of s-polarized (light polarized parallel to the nanoslit long axis) and p-polarized (light is polarized perpendicular the nanoslit long axis) light for the sample with thickness 100 nm.

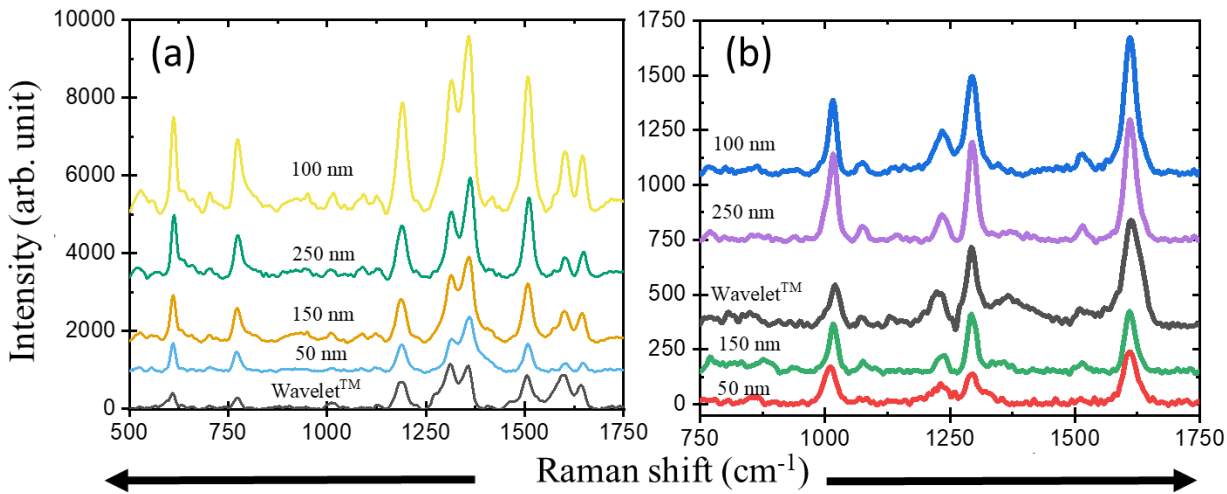

**Figure 4.** The average SERS spectra of (a) Rh6G and (b) BPY on AgNS samples. The spectra have been plotted by offset in the increasing order of their SERS intensity for clarity.

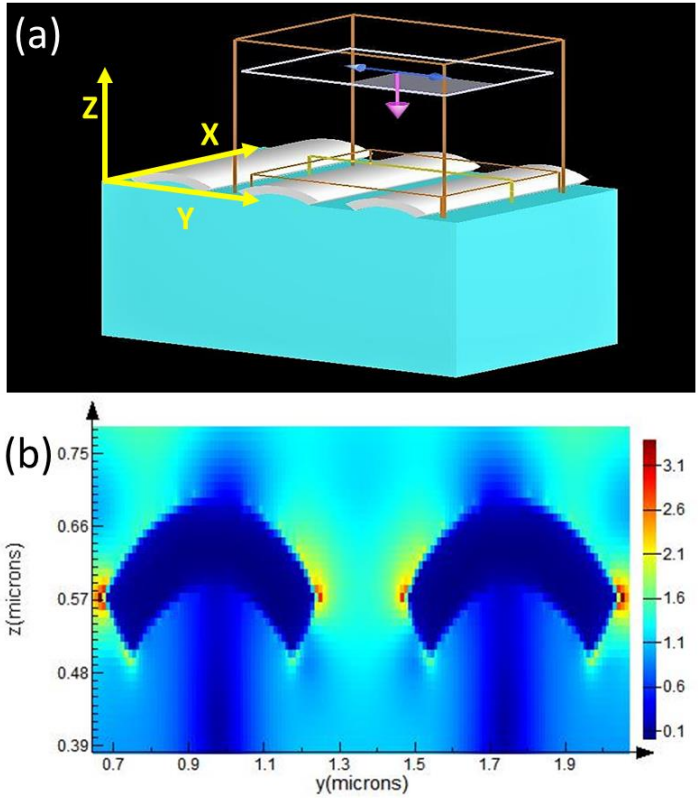

Figure 5 (a) The structure used for our simulation. The periodic boundary condition is employed in both x and y- direction to extend the geometry to infinite. Perfectly matched layers (PMLs) are used to absorb field components at the grid edges in the ± z-directions; (b) The local field enhancement for the sample S4.

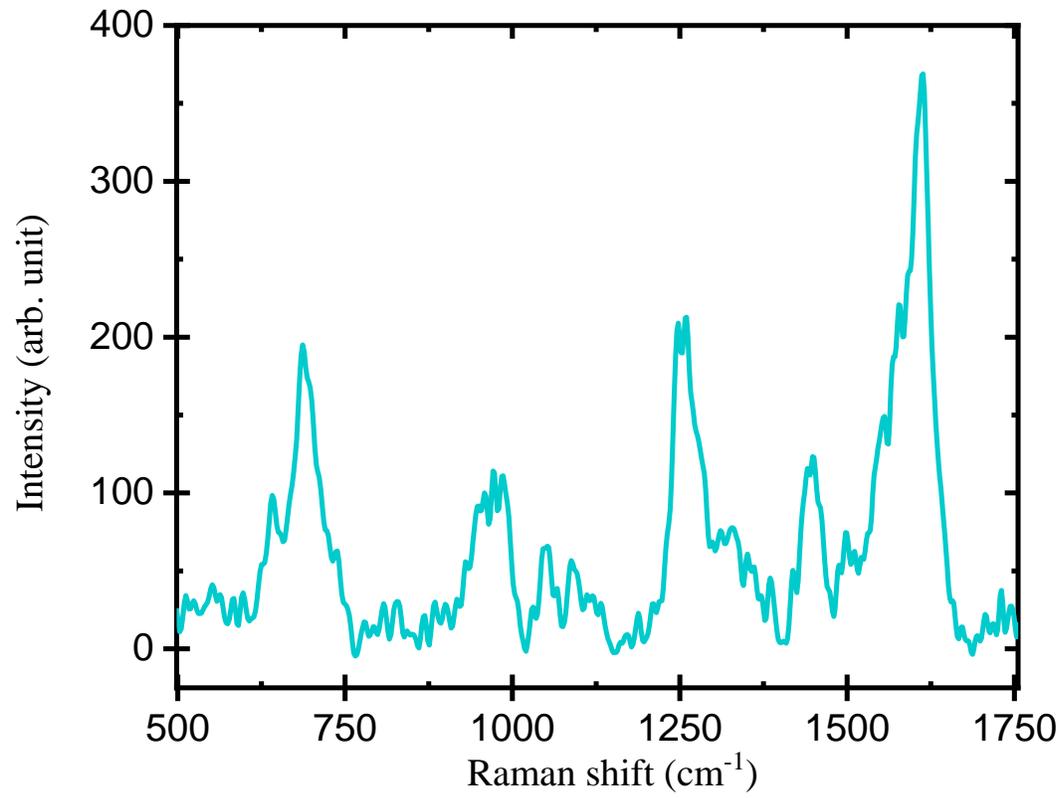

Figure 6. The SERS spectra of .06 mg/ml bilirubin solution on AgNS sample with thickness 100 nm.

# Table

**Table 1**. Calculated height and roughness from AFM images; slit width, and film width from SEM images, and calculated height

| Sample | QCM Thickness, $t$ (nm) | Film thickness, $h$ (nm) (calculated) | Slit width, $s$ (nm) | $h/s$ | LSPR (nm) | SPP (nm) | Roughness | $h$ (from AFM) |
|---|---|---|---|---|---|---|---|---|
| S2 | 206 | 50 | 170.6 | 0.29 | 720 | 754 | 39.2 | 53.5 |
| S4 | 413 | 100 | 149.3 | 0.67 | 713 | 760 | 42.5 | 97.4 |
| S6 | 620 | 150 | 115.5 | 1.3 | 704 | 767 | 39.6 | 148.2 |
| S10 | 1033 | 250 | 74.5 | 3.36 | 685 | 790 | 34.2 | 260.1 |